# Preliminary Analysis of a Potential Asteroid from the IASC Campaign Using Pan-STARRS Data

Dharanesh Palaniappan

May 2025

## ABSTRACT

In this paper, we present a detailed analysis of a potential asteroid candidate, ISP0010, using images provided by the Pan-STARRS 1 telescope as part of our participation in the International Astronomical Search Collaboration campaign. IASC is one of the leading contributors to the detection of various Near-Earth Objects and Main Belt asteroids. Using Astrometrica software we analysed those images and extracted their key parameters such as signal-to-noise ratio, right ascension, declination, flux, FWHM, RMS, and magnitude. Later, we analysed its trajectory, apparent angular speed, and error analyses using python and its libraries. In the end, we concluded that ISP0010 would be a potential candidate for an asteroid. However, on cross-verifying with the Minor Planet Centre database, we found that ISP0010 was already reported on February 27, 2021, and a total of 25 observations has been reported up to the date and currently the object has been designated as 2021 CP66. Our date of observation of 2021 CP66 was February 27, 2025. This study thus provides an independent confirmation and characterization of 2021 CP66, demonstrating the accuracy of our analysis and underscoring the importance of thorough database verification in asteroid detection.

## 1. Introduction

When the sun formed, the remaining dust and gas clouds clumped together to form the planets we see now, but not all the materials from the molecular cloud became full planets; some of them formed into some planetesimals like asteroids, comets and dwarf planets. Asteroids are mostly small rocky objects that orbit our sun in an elliptical orbit. These asteroids and dwarf planets are also called minor planets since they share some similar characteristics to a planet. In the early days, some planetesimals like Pluto and Ceres were considered as one of the planets in our solar system but as we study more about these planetesimals, IAU put some criteria for an object to be considered as a planet. It should be nearly spherical, orbit the sun and should clear orbit of other bodies. Ceres and Pluto followed the first two criteria, but they failed the third criteria by sharing their orbit with thousands of objects.

  Generally, the sizes of asteroids range from a few meters to several kilometres. The largest asteroid is Ceres, about ~933 km in diameter, which is classified as a dwarf planet, and Vesta, about ~530 km is the largest asteroid that does not fall under a dwarf planet. The space between Mars and Jupiter holds most of the asteroids in our solar system. The asteroids that orbit the sun between Mars and Jupiter are called Main Belt Asteroids. Asteroids that are found at the

Lagrange points of the planets are called Trojan Asteroids. And some asteroids orbit within the inner solar system.

Most of the asteroids are formed at the same time planets are formed and they are unchanged much, so by studying and understanding those asteroids we can get meaningful information of our early solar systems, revealing what things were like when the planets were forming. Moreover, monitoring asteroids especially NEO is important in assessing the possibility of impact of asteroids on Earth.

## 2. Minor Planet Detection: Insights from IASC and Pan-STARRS

### 2.1  *IASC overview*

The International Astronomical Search Collaboration (IASC) is a global citizen scientist programme that provides high-quality astronomical data to participants around the world. IASC is the leading contributor in identifying many main-belt asteroids and near-Earth objects using the images from Pan-STARRS. Through IASC, these citizen scientists can make their original astronomical discoveries and gain knowledge of hands-on astronomical data analysis. We have participated in the 2025 campaign from February 21 to March 19.

### 2.2 Pan-*STARRS and Data Collection*

The Panoramic Survey Telescope and Rapid Response System is one of the renowned astronomical observatories all around the world. They have two major telescopes. The Pan-STARRS 1 telescope, which is situated in Maui, Hawaii and it is one of the world's leading NEO-discovering telescopes. The Pan-STARRS 2 telescope, is also situated on the same Maui Island. The images captured by these two telescopes are collected by the IASC, and they send these images to the participants as multiple image sets. They analyse these images using Astrometrica software and check for true signatures such as constant motion in a straight line across all frames, SNR value and consistency of magnitude to find a potential candidate for an asteroid.

## 3. Asteroid Detection and Data Analysis

### 3.1 *Astrometric and Photometric Analysis*

We received an image set containing 4 images, which were taken at a regular interval of about a 17-minute gap between each image on February 27, 2025. These images were analysed using Astrometrica, a software that helps in astrometric and photometric analysis and is used to detect asteroids and perform data analysis. The software will calibrate and align the images we gave with reference star catalogues to ensure accurate positioning. By blinking the set of 4 images, we identified moving objects showing consistent motion across all frames. Astrometrica then automatically measured key parameters such as right ascension, declination, signal-to-noise ratio, flux, magnitude, and image quality indicators (e.g., FWHM and RMS). Those parameters are given below in the following Table 3.1.

**Parameters obtained from the astrometic and photometric analysis of 2021 CP66**

| Analysis | RA | Dec | RMS | SNR | Flux | G | FWHM | Time and Date |
|---|---|---|---|---|---|---|---|---|
| Image 1 | 12 33 23.586 | -08 02 44.38 | 0.297 | 3.4 | 573 | 21.6 | 0.3" | 11:48:01.7UT |
| Image 2 | 12 33 23.211 | -08 02 42.95 | 0.181 | 4.1 | 655 | 21.6 | 0.5" | 12:05:11.3UT |
| Image 3 | 12 33 22.816 | - 08 02 41.64 | 0.135 | 4.9 | 905 | 21.2 | 0.7" | 12:22:23.2UT |
| Image 4 | 12 33 22.424 | -08 02 40.60 | 0.187 | 4.5 | 1264 | 21.3 | 0.9" | 12:39:36.1UT |

RA – Right Ascension; Dec – Declination; RMS – Root Mean Square Error; SNR – Signal-to-Noise Ratio, Flux – Total photon count received; G – Apparent Magnitude; FWHM – Full Width at Half Maximum

**Table 3.1**

After obtaining these parameters using Astrometrica, we used Python and its libraries to study the objects other few characteristics that further strengthen the possibility of being an asteroid.

### 3.2 *Analysis of the Asteroid*

### 3.2.1 *Angular speed*

We all know that using this simple equation we can find the angular speed.

**Angular speed = Angular distance / Time**

$$\omega = \frac{\theta}{t}$$

Using RA, Dec and Time of the image, we can calculate the angular speed of the object. For that, first we need to calculate the angular separation of the object across each frame. So, we need to use another formula.

$$\theta \approx \sqrt{(\Delta\alpha.\cos\delta)^2 + (\Delta\delta)^2}$$ (Where the α is Right Ascension and δ is Declination)

This formula for angular distance was taken from Astrophysics: A Review of Coordinates by Prof. Michael Richmond, published on LibreTexts.

On substituting these RA, Dec and time from Table 3.1 in the above equation we can get the angular distance of each image.

Images (1 and 2) θ ≈ **5.75028 arcseconds**; t = **0.286 hours**; ω ≈ **20.10587 arcseconds/ hour**
Images (2 and 3) θ ≈ **6.01111 arcseconds**; t = **0.287 hours**; ω ≈ **20.97121 arcseconds/ hour**
Images (3 and 4) θ ≈ **5.91429 arcseconds**; t = **0.285 hours**; ω ≈ **20.61327 arcseconds/ hour**

The angular velocity of the asteroid across those 4 images is ≈ **20.56345 arcseconds/ hour**.

As mentioned in the ISO Deep Asteroid Survey (Tedesco, 2001, p. 5). The apparent rate of motion of MBA is 0-60 arcsec/hour, so from this it can be inferred our object is an MBA.

(Although **arcseconds/ second** is the standard unit in the high-precision astrometry, **arcseconds/ hour** was chosen here for ease of interpretation, and Python reduced the possibility of approximation error.)

### 3.2.2 *RA and De velocity*

The RA velocity talks about the velocity of the object moving on the East-West axis. While the De velocity talks about the velocity of the object moving along the North-South axis. Let us calculate the RA and De velocity using the equations given below,

$$\text{RA velocity} = \frac{\Delta\alpha \cdot \cos(\delta)}{\Delta t}$$

$$\text{De Velocity} = \frac{\Delta\delta}{\Delta t}$$

This formula for RA and Dec velocity was obtained from formula for angular distance from Astrophysics: A Review of Coordinates by Prof. Michael Richmond, published on LibreTexts.
For image (1 and 2) $V_{ra}$ ≈ **-19.47424 arcseconds/ hour**; $V_{de}$ ≈ **5.0000 arcseconds/ hour**
For image (2 and 3) $V_{ra}$ ≈ **-20.46717 arcseconds/ hour**; $V_{de}$ ≈ **4.57021 arcseconds/ hour**
For image (3 and 4) $V_{ra}$ ≈ **-20.29207 arcseconds/ hour**; $V_{de}$ ≈ **3.62474 arcseconds/ hour**

The Right Asencion velocity ≈ − **20.07783 arcseconds/hour**
The Declination Velocity    ≈    **4.39831 arcseconds/hour**

Based on the values of the RA and Dec, it can be inferred that the 2021 CP66 is moving in **North-West direction.**

### 3.2.3 *Trajectory of the asteroid*

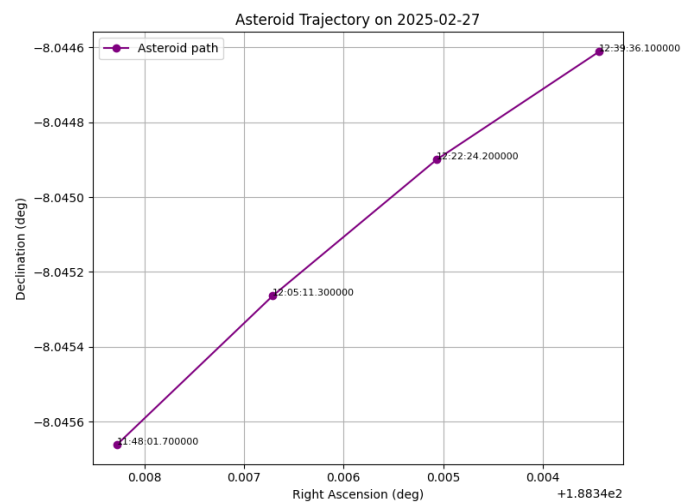

**Figure 3.1**

By plotting the graph for Right Ascension and Declination, we can see the trajectory of the object moving across the celestial sphere. It also helps us to predict the direction of the object and can predict the coordinates of an asteroid in the future. Figure 3.2 shows the trajectory of 2021 CP66.

For an asteroid with limited data of 4 images, the graph will appear to be linear. The graph we obtained also appears to be nearly linear but with a slight curve which indicates the asteroid is moving in its orbital path or it may be also due to the gravitational influence of nearby objects. But for an asteroid with multiple data, the graph will show clear curvature of its orbital path, and we can also study its rotation period. This graph gives meaningful insights about why our object could be an asteroid but with limited data we could not further proceed detailed analysis.

### 3.2.4 *Analysis of Photometric and Observational Parameters*

In Figure 3.2, we see the photometric analysis of the object. In this, the Flux vs Time graph shows that the brightness of the object, which increases from 573 to 1264, and the G value decreases from 21.6 to 21.3. This pattern suggests that the asteroid is getting brighter over time, which might be due to the rotation of the asteroid, change of phase angle or varying distance from the observer. As expected, the SNR value also increases. Even though the SNR values are less than 5, the motion of the object over the 4 images is constant, at a regular pace and moves in a straight path, this indicates that there is less possibility of the object being an artifact or noise.

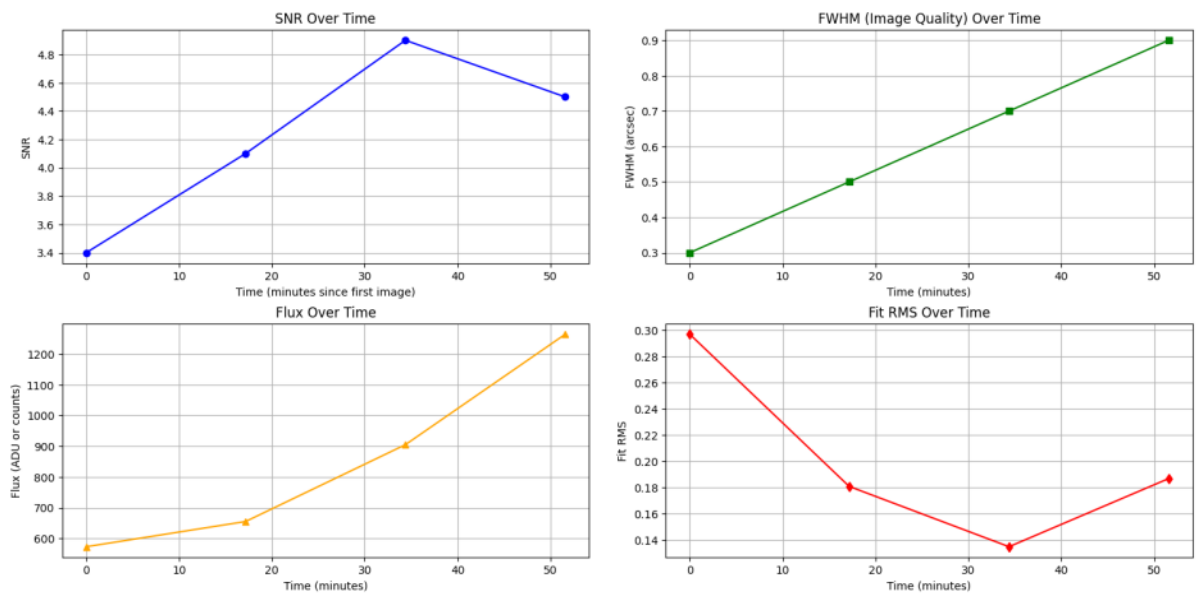

**Figure 3.2**

The RMS values are 0.297, 0.181, 0.135, and 0.187, which show very good Fit RMS values. The FWHM values are increasing, which suggests that there may be atmospheric turbulence or error in focus, The readout noise ranges from 5.5 to 5.95, which is slightly above the threshold value it is suggesting that there might be slight uncertainty. Also, the airmass values (1.14 to 1.18) are good, but there might be some atmospheric turbulence or absorption of the light. The analysis is limited because we only have 4 images.

The observed readings indicate that it will be most likely an asteroid and with limited reading, we could not infer other physical features like albedo, rotation period and other features.

### 3.2.5 *Noise analysis*

Noise analysis is important for any astronomical observation. A good observation should have less noise and a strong signal. Let us calculate our DFT SNR.

Flux = 573, 655, 905, 1264; SNR = 3.4, 4.1, 4.9, 4.5

$Noise = \frac{Flux}{SNR}$ = 68.53, 159.76, 184.69, 280.89

Let us calculate our DFT SNR.

$$x[k] = \sum_{n=0}^{N-1} x[n] e^{-\frac{i2\pi kn}{N}}$$ (Discrete Fourier Transform)

N = Number of images; k = Frequency component (Let us keep it as 1)

Using the method outlined in the NumXL article "Calculating Signal-to-Noise Ratio Using DFT", we derived the DFT-based SNR calculation for this study.

Using Euler's identities,

$$X[1] = x[0] + x[1]e^{-i\pi/2} + x[2]e^{-i\pi} + x[3]e^{-i3\pi/2}$$

Substituting our Noise values in this equation, we get

X [1] = -332- 609i

Now let us calculate amplitude,

$A_a$ = [(-322)$^2$ + (609)$^2$]$^{0.5}$ = 693.62

Similarly calculate that with noise values

$$X1 = 168.53 + 159.76(-i) + 184.69(-1) + 280.89(i)$$

$A_b$ = [(-16.16)$^2$ + (121.13)$^2$]$^{0.5}$

$A_b$ = 122.21

Fourier Amplitude of Signal ($A_1$): ≈ 693.62; Fourier Amplitude of Noise ($A_1$): ≈ 122.21

By dividing $A_a$ and $A_b$, we can the SNR by DFT,

SNR = $\frac{693.62}{122.21}$ = 5.68

Calculated SNR (using Fourier amplitudes): ≈ 5.68

Usually, an SNR value above 5 is generally considered a good SNR value and most likely, it won't be noise or artifact. Our value is 5.68, which suggests we have good signal compared to noise. With the limitation of 4 images, DFT method might be less sensitive compared to the

SNR we obtained through photometric analysis using Astrometrica. But this DFT SNR will be reliable in analysing basic parameters.

## 4. Result

In this study, we analysed a preliminary asteroid candidate P127jP7 using four FITS images provided by the International Astronomical Search Collaboration (IASC) from the Pan-STARRS telescope. Using Astrometrica software, we analysed of the FITS images, the object shown constant motion across all images. we obtained a few key parameters such as Right Ascension (RA), Declination (Dec), RMS error, Signal-to-Noise Ratio (SNR), flux, apparent magnitude (G), and Full Width at Half Maximum (FWHM). With these data, we found that the Angular velocity of the object, which is approximately **20.61327 arcseconds/ hour.**

## 5. Conclusion

Such a low angular speed suggests that it might be a Main belt asteroid rather than an NEO, because most NEO's have high angular speeds compared to main belt asteroids. The object's motion is along the North-West direction. Even though the SNR values taken from the analysis with Astrometrica were below the threshold, the SNR we obtained through Fourier analysis is 5.68, which adds extra proof that our object will most likely will be a real astronomical object rather than a noise. However, the analysis is limited by the small number of images and moderate SNR, limiting the ability to determine additional physical properties such as rotation period, albedo, or detailed orbital elements. During the initial analysis of the IASC campaign, the object did not have a designation and was considered a preliminary asteroid candidate. Our preliminary analysis yielded the result that the object will be most likely an asteroid. Further observations are needed to confirm its other characteristics like orbit, true speed, and other physical features. With multiple images of the object, we can find more parameters of the asteroid.

To avoid confusion, we avoided any assumptions about the object's orbital parameters or physical characteristics. Instead, we focused on performing calculations strictly based on the data we have. This cautious approach ensures that our analysis remains grounded in evidence and avoids speculative errors.

However, a recent cross-check with the Minor Planet Center (MPC) database revealed that the object has now been officially designated as a known asteroid and it has been observed several times before and first observed on February 27, 2021. This shows the nature of asteroid discovery and the importance of multiple observations. Thus, later cross verification with the MPC database validates the accuracy of our analysis. The following Table 4.1 illustrates the findings by MPC.

| | | | |
|---|---|---|---|
| Designation | 2021 CP66 | Perihelion distance | 2.1876834 AU |
| Orbit type | Main Belt Asteroid | Aphelion distance | 2.573 AU |
| Argument of perihelion (°) | 293.25456 | Semimajor axis | 2.3801525 AU |
| Ascending node (°) | 279.38991 | Mean anomaly (°) | 347.46670 |
| Inclination (°) | 2.48017 | Absolute magnitude | 18.88 |
| Eccentricity | 0.0808642 | Phase slope | 0.15 |
| Orbital period (yrs) | 3.67 | Residual rms (arc-secs) | 0.40 |

**Table 4.1**

## 6. Acknowledgment

I sincerely would like to thank Pan-STARRS for providing those image sets and IASC for providing this wonderful opportunity for us to give hands-on data analysis experience. I also extend my gratitude to Astrometrica software for providing such a user-friendly workspace.